\documentclass[preprint]{aastex}

\shorttitle{Close binary stars.~VII}
\shortauthors{Rucinski}

\begin{document}

\title{Radial Velocity Studies of Close Binary 
Stars.~VII.\footnote{Based on the data obtained at the David Dunlap 
Observatory, University of Toronto.} \\
Methods and Uncertainties.}

\author{Slavek M. Rucinski}
\affil{David Dunlap Observatory, University of Toronto\\
P.O.Box 360, Richmond Hill, Ontario, Canada L4C 4Y6}
\email{rucinski@astro.utoronto.ca}


\begin{abstract}
Methods used in the radial-velocity program of
short-period binary systems at the David Dunlap Observatory 
are described with particular stress on the Broadening
Function (BF) formalism. This formalism has permitted 
determination of radial velocities from
complex spectra of multiple-component systems with 
component stars showing very different degree of rotational line
broadening. The statistics of random errors of orbital parameters
is discussed on the basis of the available orbital solutions 
presented in the six previous papers of the series, 
each with ten orbits. The difficult
matter of systematic uncertainties in orbital parameters is
illustrated for one typical case of GM~Dra from 
the most recent Paper~VI. 
\end{abstract}


\section{INTRODUCTION}
\label{sec1}

This paper should be considered as 
a companion and supplement to the previous papers of our series
of radial velocity studies of close binary stars:
\citet[Paper I]{ddo1}, \citet[Paper II]{ddo2},
\citet[Paper III]{ddo3}, \citet[Paper IV]{ddo4}, 
\citet[Paper V]{ddo5}, \citet[Paper VI]{ddo6}.

The current program of radial velocity observations
of close binary systems with periods shorter than one day 
is approximately at its half-way point. 
Our methods have been evolving slightly during the execution of the 60 
radial velocity orbits presented in the six papers of the
series, but appear to have stabilized now, warranting a more 
detailed documentation of the essential steps in our analysis and data 
reductions. We summarize these methods and give an overview
of the uncertainties so that the results described in the previous 
and the planned future papers of the series could be better
evaluated by readers. The discussion is limited strictly to 
methodological aspects and does not include any astrophysical 
results which will be discussed after the program is concluded.

\section{INSTRUMENTATION AND OBSERVATIONS}
\label{instr}

We observe radial velocities of close binary stars
with the 1.88m telescope of the David Dunlap Observatory 
using its medium-resolution spectrograph in the Cassegrain focus. The 
angular scale in the telescope
focus is 6 arcsec per mm. We use one of the two spectrograph slits,
300 $\mu$m or 250 $\mu$m in width, 
both fixed in the E-W orientation and both 10 mm long.  
The angular widths of 1.8 and 1.5 arcsec approximately 
match the median seeing at the DDO of 1.7 arcsec. Since we started with 
the shortest-period binaries showing the strongest rotational 
line broadening, most observations have been made with 
the 300 $\mu$m slit. 
The scale reduction of the collimator--camera combination  
is 4 times resulting in the slit image of 75 $\mu$m
or 62 $\mu$m for either of the slits. Our light detector is currently 
a thick, front-illuminated CCD chip of 1024 by 1024 pixels, 19 $\mu$m 
square. Thus, the slit images have the total widths of 3.9 or 3.3 pixels
while the FWHM widths are about 2.6 and 2.2 pixels for the
respective slits. To lower the influence of the readout noise, the 
two-dimensional CCD images are on-chip binned 4-times in the direction 
perpendicular to the dispersion direction.

Most of the spectral data have been obtained using the
1800 lines/mm diffraction grating, with the spectral window centered on
the magnesium triplet Mg~I ``b'' at 5184 \AA. For solar-type stars,
this region is very rich in spectral lines, which is an essential
consideration for our method of radial-velocity measurements --
through broadening functions -- 
to succeed. The main-sequence stars of spectral types of middle-A to 
middle-K are practically the only stars found in close 
binaries with orbital periods shorter than one day.  
At 5184 \AA, the spectrograph delivers 0.204 \AA/(19 $\mu$m pixel) 
or about 11.8 km~s$^{-1}$ per pixel. As it 
is well known, when cross-correlation or similar techniques are used, 
narrow, properly-sampled, symmetric spectral features 
can be usually measured to better than about
1/10 part of the pixel size, with the 
accuracy growing in relation to the total length of the spectrum. 
In our case, the spectrum has the length of 208 \AA\ 
so that we can rather easily 
determine velocities of sharp-line stars with accuracy of about
1~km~s$^{-1}$, as has been verified by many 
observational programs at the DDO (see the end of this section).
The accuracy for broad-lined spectra of binary components 
is obviously lower and depends on a combination of many factors. We 
discuss the random errors in Section~\ref{rand}, while systematic
uncertainties specific to close binary stars are discussed in 
Section~\ref{syst}.

The spectrograph is known to show some flexure limiting exposures to 
about 30 minutes. This has not been a real limitation in the 
observations because our program stars have short periods and in fact
require exposures to be 
no longer than 15 -- 20 minutes to prevent the radial-velocity 
smearing. We normally take comparison-lamp (FeAr) spectra before and 
after each exposure, but sometimes, for the shortest-period binaries, 
we take two or three stellar exposures for a set of bracketing
comparison spectra 
spaced by no more than half an hour. The flat field lamp is an internal 
one in the spectrograph, but we occasionally take also sky flat-field 
spectra. 

We observe typically 3 to 5 radial-velocity standard stars per night. 
These stars are selected to have spectra of similar spectral types to 
our program stars, to serve later as templates in our technique of 
radial-velocity measurements through broadening functions (see 
Sections~\ref{bf} -- \ref{measur}). An inter-comparison 
of radial velocities of standard stars gives an estimate of random
errors at the level of about $1.0 - 1.2$ km~s$^{-1}$. This agrees with
the results for Cepheids observed at the DDO by \citet{SE94} 
and \citet{eva00} where the errors were estimated at $1.0 - 1.3$
km~s$^{-1}$. A fraction in these errors may come from our continuing
use of the IAU standard-star list as published in the 
Nautical Almanac 1995. As explained in \citet{ste99}, the  
IAU list contained at that time a few stars which are unsuitable as
radial velocity standards. 
Since we used many different standard stars from the
IAU list, these systematic errors averaged out 
to some degree and manifested themselves mostly as random errors.
We now use exclusively the list of \citet{ste99}, but the
results of our program may be affected by the uncertainties in the
old standard velocity data at a level of 0.2 -- 0.5 km~s$^{-1}$.

\section{INITIAL ANALYSIS OF SPECTRA}
\label{init}

The reductions consist of several stages. Stage \#1 consists of a 
transformation from two-dimensional images to one-dimensional,
wavelength-calibrated, rectified spectra. All the steps, 
starting with de-biasing and flat-fielding, utilize the standard 
techniques within IRAF\footnote{IRAF is distributed by the National 
Optical Astronomy Observatories, 
which are operated by the Association of
Universities for Research in Astronomy, Inc., under cooperative
agreement with the National Science Foundation.}. 
We make sure to use a consistent set of low-order polynomials for 
the dispersion relation and use the IRAF rejection algorithm for
rectification of the spectra; both steps are 
facilitated by the short length of our spectra
within which the dispersion and the CCD sensitivity vary only slightly 
and in a smooth way. 

Figure~\ref{fig1} shows a typical spectrum for our program of the
binary KR~Com, one of 
the stars presented in the most recent paper of the
series \citep[Paper VI]{ddo6}. It is a typical, yet difficult case, 
in the sense that we have been frequently dealing 
with rather complex, multi-component 
spectra even among relatively bright
stars (7th magnitude in this case). 
Some systems of our program have been 
previously known binaries, too difficult to handle using traditional 
(including cross-correlation) methods, but many were recently
discovered as photometric variables. 

\placefigure{fig1}

The spectra of KR~Com are 
dominated by the third, slowly rotating component which -- although 
the fainter one in the visual system -- dominates the spectral
appearance and produces sharp, easily
identifiable spectral lines. The triplicity of KR~Com 
went apparently unnoticed for so long mostly because the spectra,
superficially, look like those of a single, slowly-rotating
star and -- paradoxically -- the spectrum of the brighter binary component
is not normally visible. The presence of the binary, which produces the
broad spectral signature, manifests itself spectrally only through
merging of more common, weaker lines and the general lowering 
of the continuum and is difficult to notice in low S/N spectra.
The low-level photometric variability of the whole system is due to the 
contact binary which is the brighter component in the system,
but the variability signal is sufficiently ``diluted'' in the 
combined light of the system that it took the high 
quality of the Hipparcos photometry to discover it.  

Spectra like that shown in the left panel of 
Figure~\ref{fig1} are not analyzed directly,
but are subject to the broadening-function extraction process, which is
followed by measurements of radial velocities.

\section{WHY WE DO NOT USE THE CROSS CORRELATION FUNCTION}
\label{ccf}

Step \#2 of the analysis is determination of the broadening 
function (BF). The BF approach was described before in 
\citet{ruc92,ruc99} and is discussed more extensively in 
Section~\ref{bf}. In essence, it consists of a linear, least-squares 
determination of the broadening convolution kernel from 
rotationally- and orbitally-broadened spectra,
utilizing spectra of sharp-line, slowly rotating 
radial-velocity standards. We do not use the popular cross-correlation 
function (CCF) technique because it appears to give inferior and biased
results for close binary systems. We now try to explain this
rather strong statement.

\begin{enumerate}
\item The CCF {\it combines\/} the broadening 
of the program spectrum with
that of the template, with the resulting loss of resolution,
while the BF approach attempts to {\it remove\/} the common 
broadening contributions. Only if the template spectrum were
a series of delta-functions, would the results be the same.
\item The definition of the baseline in the CCF is usually
difficult and may lead to problems when relative luminosities of
components are determined.
\item Outside of the main peak which is used for
radial velocity determination, 
the CCF always shows a fringing pattern
which may affect the strength and intensity of secondary
correlation peaks for multiple systems. For very close
binaries, the secondary fringes frequently produce
the ``peak-pulling'' effect of the systematically smaller 
radial-velocity amplitudes, but a more complex interaction is
entirely possible.
\item The shape of the CCF beyond the main correlation
peak depends on the shape of the stellar 
spectrum. {\it For the same star, observations in different
parts of the spectrum define different CCF's.}
This problem is rarely recognized and is particularly severe
for sparse spectra, when the CCF is analyzed 
over a wide range of correlation lags.  
\end{enumerate}

The problems listed above are illustrated in Figure~\ref{fig1}
for the case of the triple system KR~Com.
In the right panel, we show a comparison of the BF with the CCF
for the same spectra. 
While the BF very clearly shows all three components in the
system, it would be very difficult separate the three signatures
using the CCF. The superior resolution offered by 
the BF approach has permitted
us to analyze spectra with very strong rotational broadening,
combined with situations of three or more sets
of blended lines in triple and quadruple systems, with component stars
showing different amounts of rotational broadening. Such systems
have been frequently abandoned in the past because of insurmountable
difficulties with separating and measuring radial velocities
of individual components. 

The problems of the baseline location and fringing in the CCF, 
as well as of 
the dependence on the shape of the stellar spectrum are illustrated in
Figure~\ref{fig2}. This figure contains a result of the following
data-processing experiment. A high quality, but somewhat sparse 
stellar spectrum (left panel, 
sampled at equal velocity steps of 0.88 km~s$^{-1}$)
was convolved with the single-star rotational-broadening pattern with 
$V \sin i = 88$ km~s$^{-1}$ and then subject to the CCF
determination. No noise was added, so the CCF is basically perfectly
determined and can be used for a direct comparison with the assumed broadening
function. The right panel of the figure compares the
assumed broadening profile (dotted line), which is 
the same as the BF, with the CCF (full line). 
The strong fringes in the CCF are very well visible. 
In this particular case, the unusual strength 
the positive fringes results from the low density of spectral
features in the original spectrum and illustrates the dependence of the CCF
on the stellar spectrum. While the BF formalism is insensitive
to the density and distribution of the spectral lines, 
the CCF -- beyond the main peak -- does depend on the spectral region.
Thus, sparse spectra will lead to less-well defined broadening functions
with larger {\it random errors\/} (simply because of the lower 
information content), while the CCF will additionally show {\it systematic 
differences\/} in the fringing pattern outside the main peak.  

The negative fringes which are always present in the CCF
can produce a quasi-baseline around the main correlation peak
at a very different level than expected. In the
case shown in Figure~\ref{fig2}, the local baseline
is located about $-0.1$ below the originally assumed broadening
profile. If only a small part of the CCF were used, this is where
the baseline would normally be located. Since a CCF
would rarely be used for anything else but a 
radial-velocity determination from
the correlation peak, the exact location of the baseline 
may seem immaterial. However, when a secondary star is added
to the picture, with the similar rotational broadening and 
a velocity separation comparable to the rotational broadening,
as is the case for short-period binary stars, then the secondary 
peak in the CCF will definitely interact with
the primary-star fringing pattern; there will be also the
reverse interaction of the secondary pattern with the primary
peak. For single objects, the fringes are basically irrelevant, 
so that very close or identical radial velocities are
determined using both techniques. 
Problems occur when multiple components are present
in the spectra and are particularly severe in the CCF when broadening
of the lines is of the same scale as the line splitting,
exactly the situation we face in our program.

We recognize that a method based on the two-dimensional 
cross-correlation function 
called TODCOR \citep{todcor1,todcor3} has 
been developed and successfully applied to several multi-lined 
stellar systems showing sharp spectral lines. 
We did not attempt to use this technique mostly because we 
feel more comfortable with a tool developed by ourselves, but also 
because (1)~TODCOR is designed for sharp-lined spectra
and has not been demonstrated to work for the 
very broad lines of contact binaries, and 
(2)~we frequently deal with mixed very broad and narrow 
spectral signatures which would require extension of the
TODCOR capabilities even further. 
We note that the non-linear nature of the 
cross-correlation complicates derivation of relative luminosities of 
components and requires a complex calibration while our linear approach 
gives directly the relative luminosities through integration of the 
individual features in the broadening functions. 
This is particularly convenient for systems with 
components showing very different degrees of rotational broadening.

\section{BROADENING FUNCTIONS}
\label{bf}

We define the broadening function\footnote{A full 
description of the concept
of the broadening functions with examples and detailed 
programming suggestions is available at the WWW site
http://www.astro.utoronto.ca/$\sim$rucinski.}  
as a function that transforms a sharp-line spectrum of 
a standard star into a broadened spectrum of a binary or, for that
matter, of any other star showing geometrical, 
Doppler-effect line broadening. This way 
we not only determine the broadening-function shape, but also 
automatically relate absolute velocities of program stars to 
radial-velocity standards used as templates, a common 
advantage with the CCF approach. We do not use model spectra, e.g.\ 
through representation of spectral lines by delta functions. While 
the broadening functions determined that way would be cleaner and
much better defined than those utilizing 
standard-star templates, the advantage of the automatic 
relative radial-velocity calibration would be lost.

We perform all radial velocity determinations in the geocentric 
system and only later transform the results 
to the barycentric (heliocentric with 
planetary corrections) system. Thus, 
we start with a raw template spectrum 
$S_t$, with its wavelength scale in $W_t$, 
and a raw program spectrum $S_p$, with its wavelength scale 
in $W_p$. Both $S_t$ and $S_p$ are rectified and normalized to 
unity. To diminish importance of the one-to-zero discontinuities at the 
ends of the spectra, we invert them so that the absorption lines are 
represented by positive spikes: $S'_t = 1-S_t$ and $S'_p = 1-S_p$.

The spectra must be of similar spectral type. We normally use 
the templates with spectra within one spectral type. However, we have
found that
F-type templates will work reasonably well for radial velocity
determinations between middle A-types to early K-type stars; however,
the relative luminosity estimates from the individual peaks
will then be wrong. 

The spectra must be first re-sampled into equal steps in velocity. In our 
case the velocity step is typically $\Delta v=11.8$ km~s$^{-1}$. 
An auxiliary vector of wavelengths is now created
with elements: $W=W_0*(1+r)^i$, 
where $i = 0, \ldots, n-1$ is the index in the new vector and
$r=\Delta v/c$, where $c$ is the velocity of light. 
The origin of this vector, $W_0$, is selected to fall
just above both origins of $W_t$ and $W_p$ for a meaningful
interpolation of both spectra into the new wavelength scale. 
The length of $\vec{W}$ is in our case usually selected to be
$n=1000 - 1020$ spectral 
elements. The spectra $S'_t$ and $S'_p$ are linearly 
interpolated using $W$, by treating $W_t$ and $W_p$ 
as the respective abscissas, to create the spectra used
in the BF derivation: for the template, $T$, and
for the program star, $P$. 
After this is accomplished, the three wavelength vectors 
$W$, $W_t$ and $W_p$ are no longer needed because the program
and the template spectra are now in the same (geocentric) 
velocity system. We can think about them as functions $P(n)$ and $T(n)$
with the same velocity axis or vectors $\vec{P}$ and $\vec{T}$ over the
same range of indices.

The convolution operation which maps a sharp-line spectrum $T$
into a broad and/or binary-star spectrum $P$,
\begin{equation}
\label{eq1}
P(\lambda ') = \int B(\lambda ' - \lambda) \: T(\lambda) \: d \lambda
\end{equation}
can be written as an array operation,
\begin{equation}
\label{eq2}
\vec{P} = \mathbf{D} \, \vec{B}
\end{equation}
in which the rectangular array $\mathbf{D}$ is created from
the vector $\vec{T}$ by placing it as columns of $\mathbf{D}$ after
shifting it downward by one index for each successive column
(see below or, for further details, consult \citet{ruc92,ruc99}).
The broadening function is represented by a vector 
of the unknowns in the solution, $\vec{B}$.
The array $\mathbf{D}$ has the short dimension $m$ and the long
dimension $n-m+1$; it accomplishes the mapping of $T \rightarrow P$.
We normally use the odd number for the size of the broadening
function, $m$, to have it centered at the pixel symmetrically
distant from both ends. Also, for proper handling of the
ends, $m'={\rm integer}(m/2)$ 
points are removed from both ends of $\vec{P}$.

The convolution operation equivalent to Eq.~(\ref{eq2}), 
which is used in the least-squares determination of $\vec{B}$,
can be written as a system of over-determined linear equations:
\begin{equation}
\label{eq3}
P_i=\sum_{j=0}^{m-1} T_{i+m-j} \, B_j \qquad\mbox{with}\qquad
i=m', \ldots, n-m'-1
\end{equation}
The number of equations should be several times larger than the
number of unknowns, $n-m+1 > m$.
In our program, we normally use $n=1000-1020$ and $m=121$.
The size of the broadening function, $m$, 
translates into the relative velocity range (program minus template)
of $\pm 708$ km~s$^{-1}$
insuring a good definition of the BF itself and of the flat 
baseline around it. The actual size of the broadening function
is a matter of choice; sometimes we repeat the BF 
determination with a smaller $m$ for binary systems with moderate
line splitting when a wide window of over 1400 km~s$^{-1}$ is not needed. 
The point is to use as short a BF as possible because 
the quality of the determination (over-determinacy) 
increases in relation to how many times the spectra are 
longer than the BF.

Solving of the broadening function $B_j$ is 
accomplished by least-squares.  
We are strong advocates of the Singular Value Decomposition (SVD) 
technique which is particularly useful in eliminating those parts of 
the spectra that carry no information (the inter-line continuum), but 
create linear dependencies. The approach involving rejection of
small singular values is the best for restoration of the shape of the BF 
for its subsequent modeling. However, with the radial velocities in mind, 
we do not in fact eliminate any singular 
values. In this respect, we have made some departure from the original 
philosophy, but this departure has a reason; if some basis functions are 
eliminated, there exists a possibility that
the spectral features may acquire asymmetries
through an unwanted conspiracy of the basis functions which
remain in the definition of the BF. By retaining all singular values, 
we treat each element of $B_j$ (Eqs.\ (\ref{eq2}) -- (\ref{eq3}))
as a totally independent variable, 
not related in any way to its neighbors. Thus, any least-squares 
technique can be used at this stage, although we continue to use the 
SVD as easier to use and more transparent for the matrix inversion.

The details of the SVD approach to solve the array equation
Eq.~(\ref{eq2}) or its equivalent Eq.~(\ref{eq3})
for the BF vector $\vec{B}$ are described in \citet{ruc92}
and the programming examples are given in \citet{ruc99}.
Even without elimination of any singular values in the SVD
solution, this approach has an advantage that one decomposition
of the template-spectrum array,
$\mathbf{D} = \mathbf{U} \, \mathbf{W} \, \mathbf{V^T}$,
can serve to determine $\vec{B}$ from several program spectra 
through the inverted relation,
$\vec{B} = \mathbf{V} \, \mathbf{W^{-1}} \, (\mathbf{U^T}
\, \vec{P})$. 
For an excellent exposition of the SVD
technique stressing its beneficial properties, see \citet{numrec}.

Irrespective of which method of the least-square solution 
is used, the resulting
broadening functions are always very noisy and 
cannot be used for radial-velocity measurements. The reason for the
excessive noise is that each element of the
solution $B_j$ is unrelated to its neighbors and is treated as 
a separate unknown. We know, however, that 
our spectral resolution is controlled by the spectrograph slit
which introduces coupling between neighboring points of the BF.
In our case, the intrinsic smoothing introduced by one of the 
entrance slits is characterized by the FWHM of about 2.6 or 2.2 pixels. 
It is therefore reasonable to apply some smoothing to the noisy BF's. 
Superficially, this step does the same to the final shape of the BF 
as smoothing through rejection of noise absorbed 
by high-order singular values in the SVD technique; however, this 
operation is strictly local whereas removal of some
singular values may introduce non-local effects. 
Usually, we smooth the broadening functions by convolving them
with a Gaussian with $\sigma=1.5$ pixels (FWHM=3.53 pixels); 
for poor spectra of very faint stars we are
sometimes forced to use $\sigma=2.0$ (FWHM=4.71 pixels).
Such smoothing is slightly stronger than its instrumental 
counterpart by the spectrograph slit, but 
is nevertheless very small, when compared with widths of 
lines in binary stars with periods shorter than one day.

\section{RADIAL VELOCITY MEASUREMENTS}
\label{measur}

Step \#3 is the radial velocity determination 
from the broadening functions. We determine the radial velocities 
of each binary component in the geocentric system, relative to the
template star, $v_i$. Following that, the relative velocities 
are transformed to the solar system barycenter with
$V_i = v_i + (HC_p - HC_t) + V_t$, where $HC$ 
are the barycentric (sometimes called
``heliocentric'') velocity corrections resulting from the 
orbital Earth motion for the program and template spectra 
and $V_t$ is the barycentric velocity of the template star.

The broadening function $B(v)$ determined in the previous step
is defined at points separated by equal steps in
relative geocentric velocity, in our case normally 
$\Delta v=11.8$ km~s$^{-1}$, spanning the velocity range
$-708 \le V \le +708$ km~s$^{-1}$. The velocities of
stellar components are determined by simultaneous 
fitting of several Gaussian curves to that many 
spectral features as are seen in the BF. 
Thus, it would be a 4-parameter Gaussian fit for a single star 
(baseline $a_0$, strength $a_1$, position $a_2$, width $a_3$), 
a 7-parameter fit for a binary or 
a 10-parameter fit for a triple system, etc., as in:
\begin{equation}
\label{Gauss}
B(v) \simeq a_0 + \sum_{i=1}^n a_{1i} 
   \exp\left( - \left( (v-a_{2i})/a_{3i} \right) ^2 \right) 
\end{equation}
where $n$ represents the number of stellar components in the system.
We found that least-squares Gaussian fits for single stars are
usually stable, while those involving more components (binary,
triple and higher multiplicity systems) are numerically 
unstable forcing us to fix or manually adjust the width
parameters, $a_{3i}$. 

In triple systems which we encountered so far, the most typical
combination has been a broad-lined close binary 
accompanied by a sharp-lined, slowly-rotating star. In such situations, 
we first leave the width and position of the third, sharp component 
floating in order to determine best possible 
parameters for its subsequent 
subtraction from the BF. For the example shown in Figure~\ref{fig1},
the Gaussian widths for the binary components
were {\it assumed\/} at $a_{31}=110$ km~s$^{-1}$ and 
$a_{32}=70$ km~s$^{-1}$ while the width $a_{33}$
was {\it determined\/} at 24.74 km~s$^{-1}$. We found that situations 
similar to that shown in the figure require a careful removal of the 
third-component signature. In order to
define the BF for the close binary the best way possible, we
cannot remove the averaged signature of the third star from
many spectra and must subtract it as it is defined for the same observation.
There may be many reasons why subtraction of the averaged 
third-star peak leaves too large residuals, including
small changes in the effective resolution, imperfections in the
geocentric to barycentric transformations or instabilities in the
spectrograph. Obviously, this approach reduces the accuracy of the
third-star velocities, but our goal has been
to determine the best velocities for the close binary so that we 
accept this limitation. The BF for the binary is usually very 
well defined, see for example Figure~4 in Paper~IV for HT~Vir.
 
The random radial-velocity errors for binaries in occurring in
triple systems are
only slightly larger than for the isolated binaries, typically
by less than 1~km~s$^{-1}$ in the errors of
$V_0$, $K_1$ and $K_2$, which simply reflects more
degrees of freedom in the problem (see the discussion 
in Section~\ref{rand} and in Figure~\ref{fig4}). Much more
difficult to characterize are systematic uncertainties. One manifestation
of such uncertainties is the occasional presence of an
undesirable ``cross-talk'' in the 
three-feature fits in the sense that the third-star velocities 
sometimes correlate with the binary phase. We always check the
third-star velocities for the dependence on the binary phase
and sometimes find a correlation.
(For a rather extreme case, see the discussion of II~UMa in Paper~VI).
It is usually quite difficult to find reasons for the cross-talk
and each case seems to be unique. Faintness of the star and/or
poor spectra certainly magnify the problem which depends on such
factors as the location of the third peak in the BF relative to the
peaks for the binary system stars (i.e.\ with which component the third peak 
mergers most of the time) or the overall degree of the line splitting 
for the binary system (which depends on the orbital inclination).
Typically, the cross-talk increases the
error per observation of the third-star from the expected level (for a
sharp-line star) of $1.2 - 1.3$~km~s$^{-1}$, to the level of 
$1.5 - 2.5$~km~s$^{-1}$. Except for noting the presence of
the cross-talk, we are not in position to study it more extensively
given the different type of the binary-phase dependence in each case. 
Our hope is that the cross-talk will average out in the velocities
of the third component, although the final proof will come only through
external comparisons. The stress has been always on the quality of
the binary solutions, perhaps at the expense of the quality of
the radial-velocity data for the third components.

We measure the radial-velocities for the binary components, and -- 
if necessary, of a spectroscopic companion -- but
do not estimate the accuracy of the radial-velocity measurements
at this stage. In principle, it is possible to 
establish a relation between 
the signal-to-noise (S/N) in the spectrum and in the broadening function
\citep{ruc93b}, 
but further propagation of the errors into the velocity errors is
more complex and depends on many factors. While such a relation 
would be definitely needed for full modeling of the BF's, we feel that 
complexity of the error analysis is not warranted in our case. 
Thus, we do not determine the radial-velocity errors from the 
individual BF's, but evaluate them externally later 
from the orbital velocity solutions. 
Such estimates may be perhaps overly pessimistic, as they 
incorporate systematic deviations from the orbital motion models.
Most importantly, however,
the random errors are not the limiting factor in
our results; the real difficulty is in evaluation of systematic
uncertainties. We address this issue in Section~\ref{syst}, after
describing the orbital solutions (Section~\ref{orb})
and the externally-evaluated random errors (Section~\ref{rand}).

\section{ORBITAL SOLUTIONS}
\label{orb}

Step \#4 of the reductions is the determination of the
radial-velocity orbit using individual velocities of both
components at all observed orbital phases. Currently, we
do so by measuring individual velocities of components,
although a more global approach involving modeling of
broadening functions would be definitely much preferable. 
The broadening functions have the potential of providing much more 
information than just simple velocity centroids so that
orbital solutions could be carried to a much higher level
of sophistication than in this series of papers.  Such use of the 
BF's was described in \citet{ruc92,ruc93a,ruc93b}, where 
modeling of the BF shape was advocated. Full modeling 
of this type requires knowledge of the orbital 
inclination, which is usually not 
available, and involves a simultaneous determination of the 
radial-velocity span, $K_1+K_2$, the mass ratio, $q$ and 
the degree-of-contact parameter, $f$. 
The complexity of such a global approach 
is the main reason why we continue to use single velocities
to characterize motions of stellar components, 
but we do recognize limitations of this approach which may 
generate systematic uncertainties in the final results;
this is discussed in Section~\ref{syst}.
We should add, that originally, this program was intended to 
provide the $V_0$ values from a small number of radial-velocity 
measurements, to relate to the then newly available Hipparcos 
tangential velocities. However with time, 
our program acquired its current 
significance as the main contributor of radial velocity orbits for 
short-period binaries (this circumstance taking place 
partly ``by default'', through a surprising lack of 
similar programs at other observatories). Thus, we continue to use the 
Gaussian fits, but recognize that all our spectra and the
broadening functions may be used for a much more extensive modeling.

All short-period binary systems observed by us
so far have circular orbits resulting in sine-curve variations of
orbital velocities. The only exception that we had
to consider is the third star in the system of HT~Vir 
(Paper~IV) on an eccentric orbit; for this case we used the 
model of \citet{mor75}. Because eclipse effects of rotationally
broadened lines change line shapes and produce undesirable 
radial-velocity shifts, we eliminate 
observations close to orbital conjunctions, usually within
the phase ranges 0.85 -- 0.15 and 0.35 -- 0.65.

The orbital solutions are obtained iteratively. First, we use 
the linear model of two sine curves and one constant value, with 
an assumed moment of the primary eclipse $T_0$. 
Thus, for $k$ observations, we 
simultaneously fit by least squares $2 \times k$ equations of the type:
\begin{eqnarray}
\label{eq5}
V_1(\phi_l)& = & V_0 \, - \, K_1 \sin\phi_l \,     + \; \; \; 0    \\
\label{eq6}
V_2(\phi_l)& = & V_0 \, - \, \;\;\;\; 0 \;\;\;\;\; \, + K_2\sin \phi_l \\
l & = & 0, \ldots, k-1 \nonumber
\end{eqnarray}
$\phi$ is the orbital phase, $\phi_l=(t_l-T_0)/P$. Similarly to $T_0$,
the period $P$ is usually taken from literature sources and is fixed; 
only in a few cases we attempted to improve its value.
The equations can be weighted at this point
when observations are of different
quality. The weighting schemes are discussed in descriptions
of stellar systems in individual papers and are given in the tables with
radial velocities. Note the sign convention in the equations, which
implies that we usually start with an assumption that the primary,
more massive component (star 1) 
is eclipsed at the photometric primary minimum.
In other words, we assume that for a contact system the configuration
is of an A-type contact binary. We identify the W-type systems when
this assumption is not valid.

The resulting $V_0$, $K_1$, $K_2$ are the first 
approximations of the orbital parameters. The next step in the
iterative solution consists of the application of the
linearized versions of Equations~(\ref{eq5}) -- (\ref{eq6}) 
for $\Delta V_0$, $\Delta K_1$, $\Delta K_2$ and 
$\Delta T_0$. We always first use any available literature value for 
$T_0$ and then improve it by solving the linearized equations
until all corrections $\Delta$ no longer change.
It is at this stage that we determine random-error
uncertainties of the orbital parameters and 
the radial-velocity errors per observation.

\section{MEAN STANDARD ERRORS}
\label{rand}

Least-squares solutions of the linearized Equations (\ref{eq5})
-- (\ref{eq6}) can provide mean standard errors of the orbital parameters
$V_0$, $K_1$, $K_2$ and $T_0$. We do not use such errors because
they usually underestimate the random error uncertainties. Instead, 
we use the ``bootstrap sampling'' technique which involves multiple
(thousands of times) re-sampling of the data with possible repetitions, 
with subsequent solutions of all such datasets. By forming statistics of 
the spread in the resulting parameters and by determining the inner
67 percent distribution ranges, we estimate equivalents of the
mean standard errors. They are sometimes close to the linear estimates, 
but usually are larger. In any case, we consider them more realistic
as they include inter-parameter correlations.

We have sufficient amount of information from all our orbital
solutions to analyze sizes and distributions of our random errors.
For that purpose, we used all the available solutions, eliminating
three systems observed at the Dominion Astrophysical Observatory,
as reported in Paper~I, and adding W~Crv described separately
\citep{WCrv}, altogether 58 orbital solutions. 

\placefigure{fig3}

The statistics of mean standard errors per single 
observation, $\epsilon_i$, for the primary ($i=1$) and the 
secondary ($i=2$) components, is shown in the first panel
of Figure~\ref{fig3}. The median values of the errors are
$\langle \epsilon_1 \rangle = 5.48$ km~s$^{-1}$ and 
$\langle \epsilon_2 \rangle = 11.50$ km~s$^{-1}$. The corresponding
distributions for the errors of the radial-velocity amplitudes, 
$\sigma (K_i)$, are shown in the upper right panel. 
The median values are $\langle \sigma (K_1) \rangle = 1.11$
km~s$^{-1}$ and $\langle \sigma (K_2) \rangle = 1.96$
 km~s$^{-1}$. The center-of-mass
velocities $V_0$ are better established than $K_i$ because
two stars contribute in each solution to one number. The
median value of these errors is
$\langle \sigma (V_0) \rangle = 1.07$ km~s$^{-1}$. Finally,
the distribution of the mean standard errors of the 
initial epoch, $\sigma (T_0)$, is shown in the last 
panel of Figure~\ref{fig3}. The median value for this error is, 
$\langle \sigma (T_0) \rangle = 0.0011$ days (about 1.5 minute).

\placefigure{fig4}

The mean standard errors of the orbital parameters 
are correlated.
The most interesting correlations are shown in Figure~\ref{fig4}. 
The two upper panels show that the mean error of the 
center-of-mass velocity, $\sigma (V_0)$, which appears to be 
a convenient measure of the quality of the orbital data.
It depends on the brightness of the system and on the
orbital period. It
is confined within $<1.5$ km~s$^{-1}$ for $V_{max} < 8.5$, but
increases to slightly over 2 km~s$^{-1}$ for $V_{max} \ge 10$ 
(the left upper panel). The scatter in $\sigma (V_0)$ increases 
for short-period systems (the right upper panel of 
Figure~\ref{fig4}), but this may be due to the fact that 
most of our targets had periods within 0.3 -- 0.6 days,
in a range where a genuine frequency maximum 
exists in the volume-limited samples of contact
binaries \citep{ruc98}. The
systems with longer periods, $P > 0.8$ days, tend to show small
errors, but these are exactly those binaries which had been
overlooked before among bright stars and have been easy 
targets for our program. The error $\sigma (V_0)$
correlates tightly with $\sigma (K_i)$ and with $\epsilon_i$, as
shown in the four lower panels of Figure~\ref{fig4}. A particularly
close correlation with the slope close to unity 
exists between $\sigma (K_1)$ and $\sigma (V_0)$ (middle left panel). 

Binaries observed in spectroscopic triple systems show slightly
larger random errors than when isolated, typically
by less than 1~km~s$^{-1}$ in the errors of $V_0$, $K_1$ and $K_2$. 
Such binaries are shown by open symbols in all plots in 
Figure~\ref{fig4}.

\section{SYSTEMATIC UNCERTAINTIES}
\label{syst}

It is difficult to evaluate systematic uncertainties of our
results. The systematic errors depend in a complex way on the orbital
parameters and couple with the random errors. 
The main source of the systematic errors is the 
measurement of radial velocities from the
broadening functions. We approximate the
center-of-mass positions with the light centroids and measure
the centroids by fitting Gaussians. 
The latter assumption, that the radial velocities of
the light centroids coincide with radial velocities of the
mass centers, is -- in general -- not fulfilled
by distorted components in close binary systems and is
particularly dangerous for contact binaries where 
the peaks in the BF's are not symmetric, with steeper outer
parts and more gently sloping inner parts. Direct modeling
of the BF's would avoid this systematic error (see the end
of this section).

Some insight into systematic uncertainties involving the
Gaussian approximation of the peaks in the broadening functions
can be obtained by applying Gaussians of various widths
and evaluating systematic shifts in the results. 
We will consider here, as a case study, of a 
typical, 9 magnitude, A-type contact system GM~Dra from
the most recent Paper~VI. 

\placefigure{fig5}

Let us first consider one broadening function for the orbital 
phase 0.283 of GM~Dra (the lowest panel of Figure~\ref{fig5}). 
For this particular broadening function,
we would normally select the best-fitting Gaussians to have the width
parameters $a_{31}=120$ km~s$^{-1}$ and $a_{32}=80$ km~s$^{-1}$ 
(see Eq.~(\ref{Gauss})). However, as an 
experiment, we considered widths between
the estimated narrowest and widest acceptable values
of $a_{3i}$: $a_{31}=100 - 140$ km~s$^{-1}$ 
and $a_{32}=60 - 100$ km~s$^{-1}$. The extreme cases are shown
by dotted and broken lines in Figure~\ref{fig5}.
For the full ranges of the widths, the change in the 
measured velocity of the primary component is
from $-28.86$ km~s$^{-1}$ to $-31.27$ km~s$^{-1}$, while
the change for the secondary component is
from $+261.44$ km~s$^{-1}$ to $+262.49$ km~s$^{-1}$. 
Thus, systematic errors in radial velocities appear to be
at a level of 1.5 to 2.5 km~s$^{-1}$,
with larger velocities (in the absolute sense) associated with
larger assumed widths of the fitting Gaussians. 

Analysis of the type presented above can be done for all available
broadening functions, at all orbital phases. The four upper
panels of Figure~\ref{fig5} show the shifts in the measured
centroids for all available observations of GM~Dra 
obtained around the orbital quadratures, within the
orbital phase ranges 0.15 -- 0.35 and 0.65 -- 0.85, as marked in the
figure. The Gaussian widths $a_{3i}$
were incremented in equal steps and for each assumed width a
full radial velocity determination was performed. As we can see
in the figure, the systematic effects are clearly present, 
especially for the secondary (less-massive) component. The shifts
are typically at the level below 2 km~s$^{-1}$ for the primary
component, but 
the shifts of the order of 5 -- 7 km~s$^{-1}$ are not uncommon 
for the secondary component. The shifts depend on side 
of the binary system (or the sign of the
radial velocity) observed at a given orbital quadrature. 
The overall tendency appears to be that the wider 
Gaussian width $a_{3i}$ result in velocities further
away from the center-of-mass velocity, i.e.\ such ones which
should lead to systematically 
larger values of the orbital amplitudes, $K_i$. This is confirmed by 
the actual determinations of the radial velocity orbits for the extreme
values of [$a_{31}$, $a_{32}$] pairs, selected to deviate from
the optimal values of 120 and 80 km~s$^{-1}$ by $\pm 20$ 
km~s$^{-1}$. The systematic changes for the
particular case of GM~Dra strongly depend on the parameter
considered. While the changes in $V_0$ are within $+0.09$ and
$-0.16$ km~s$^{-1}$, those in the amplitudes are larger:
$-0.34$ and $+0.15$ km~s$^{-1}$ for $K_1$ and as much as
$-4.30$ and $+5.97$ km~s$^{-1}$ for $K_2$. 
While the ranges of the Gaussian widths
$a_{3i}$ were intentionally exaggerated in the
experiment, to estimate the largest systematic deviations, 
we clearly see that systematic effects may set an important
limitation on our results. For comparison, 
we note that the random errors
of the orbital parameters of GM~Dra are $\sigma (V_0) = 1.52$ 
km~s$^{-1}$, $\sigma (K_1) = 1.75$ km~s$^{-1}$ and 
$\sigma (K_2) = 2.50$ km~s$^{-1}$ (Paper~VI). 
Thus, for this particular binary,
the systematic uncertainty appears to be larger than the
random error only for $K_2$, but then it is even two times
larger.

Optimally, the systematic effects resulting from the
use of different widths in the 
Gaussian fits should be evaluated for each binary 
through a process similar to that applied to GM~Dra. 
However, we feel that it is impractical
to perform similar analyses for all systems in this program.
Besides, we know that application of
the Gaussian fits is -- in any case -- a crude approximation and
that the best approach would be to {\it model\/} the broadening
functions as it was done in \citet{ruc92,ruc93a,ruc93b}. Full BF
modeling would permit inclusion of more spectra than we
utilize now because, currently, we measure for radial velocities
only those BF's which show a clear splitting of the spectral 
signatures. By addition of these spectra we would increase the 
available material by about 20 -- 30 percent, 
which would only slightly reduce random errors and
thus produce a very modest improvement in accuracy. Much
more important would be a reduction or entire 
elimination of the systematic errors,
which may reach levels of 5 -- 7 km~s$^{-1}$. For most binaries 
of this program, this would typically correspond to
about 2 -- 3 percent error in $K_i$, but in some extreme
cases of small semi-amplitudes, the errors may 
reach 10 -- 15 percent.  
While the approach involving combined radial velocity
and light-curve modeling would avoid the main systematic 
effects, it would require a considerable organizational and
computational effort, introducing large delays in 
our mostly observational program. Since our radial velocity 
observations are
-- for most systems -- the first and the only ones, we decided
to accept the level of systematic errors generated by the use
of the measuring Gaussians and make our solutions generally
available, keeping in
mind their systematic uncertainties which must be taken into
account when considering the overall accuracy of our program.

\section{CONCLUSIONS AND PLANS}
\label{plans}

The ongoing survey of close binary systems with periods shorter
than one day, currently conducted at the David Dunlap Observatory,
has resulted in a consistent set of radial velocity orbits for
sixty previously unobserved binaries to approximately 11th magnitude. 
While, at the start, the
survey concentrated on systems which simply had not been studied
before (for various reasons, but mostly because of 
inadequate instrumentation and data-analysis tools some half 
a century ago, when this field was very active),
the photometric discoveries of the Hipparcos satellite are
now dominating in numbers. There was only one Hipparcos system
among the first twenty orbits (Papers I and II), 9 such systems
among the next twenty orbits (Papers III and IV) and 15 such
systems among the most recent twenty orbits (Papers V and VI).
About 50 known, photometrically-discovered binaries still remain 
to be observed and analyzed and new ones
are constantly added to catalogs, some of them quite bright. 
Regrettably, apparently there is no similar survey for the southern
hemisphere.

Our survey is quasi-random in the sense that we observe
all short-period ($P < 1$ day), bright, previously unobserved binaries. 
With such criteria, the contact binaries absolutely dominate
in numbers. Among the 60 systems described in the previous
six papers, only 8 were
not contact systems. This is partially due to strong
selection effects against detection of detached binaries,
but mostly due to the very high frequency of contact binary systems
in the old-disk population, particularly in the period
range 0.3 to 0.5 days, but with a tail extending 
beyond one day, to about 1.3 -- 1.5 days. The high frequency of 
incidence is strongly manifested in the volume-limited OGLE sample 
and in open clusters \citep{ruc98}. Because our survey
is magnitude limited, we tend to include many
brighter systems from the tail of the distribution
between 0.5 day and our current upper limit at one
day. Otherwise, we do not
discriminate among binary systems in any other way.
In particular, the random character of the survey 
has resulted in discoveries
of the largest ($q=0.97$, V753~Mon; Paper~III) 
and the smallest ($q=0.066$, SX~Crv; Paper~V) 
known mass ratios among contact binaries.

The DDO survey is characterized by moderate random errors of 
about 1 -- 2 km~s$^{-1}$ for the orbital parameters, $V_0$,
$K_1$ and $K_2$, and -- upon completion -- can serve as a useful
database of parameters of very close binary systems. We
are aware, however, that our final parameters contain
systematic uncertainties resulting
from our radial-velocity measurement techniques. While the
use of the broadening functions permitted us to analyze
close binaries in several multiple, visual/spectroscopic systems
providing data which were too ``difficult'' before, our
extraction of individual radial velocities from the broadening 
functions, through Gaussian fitting, is a disputable approach 
for contact binary systems. Because the line-broadening for
such systems is very strong, comparable with orbital
velocities of hundreds of km~s$^{-1}$, and -- in fact --
somewhat asymmetric, our measuring technique 
may lead to systematic errors 
reaching levels of 5 -- 7 km~s$^{-1}$ or even more. 
Paradoxically, through the use of the broadening functions in
place of the cross-correlation functions, we have uncovered 
real physical reasons why the Gaussian approximation 
is only barely appropriate. 
The correct approach avoiding the systematic errors
would be to model the broadening functions and 
determine the radial velocities in terms of the mass ratio, $q$,
and the scaling factor, ($K_1+K_2$), with the shift, $V_0$. The
models would require independent input from parallel solution of
light curves, providing the orbital inclination angle,
$i$, as well as the degree-of-contact, $f$. Currently, most of the
program targets have not had their light-curves solved, and
even if some attempts have been made, we would not trust them
for the following simple reason: 
We have seen so many cases of the spectroscopic mass
ratio different from the previous photometric mass-ratio
determinations, $q_{sp} \ne q_{phot}$, that we feel very
strongly that the values of $q_{phot}$ are usually not
properly constrained and may be plainly wrong\footnote{Totally 
eclipsing systems are an exception, as pointed by
\citet{MD72a,MD72b}, but then chances of
total eclipses depend on the mass-ratio itself (a wider
range of inclinations for small values of $q$),
producing a very complex bias in the uncertainties of 
$q_{phot}$.}, leading to entirely incorrect combinations
of orbital parameters.

We envisage that the results of this survey will provide just a first
stage of an iterative process. In future, our spectroscopic 
values of mass ratio, $q_{sp}$, should permit solution of light 
curves which were previously unsolvable because to the poorly
constrained mass ratios. The derived information on ($i$, $f$) pairs 
would permit, in turn, a re-discussion of the
broadening functions and determination of the final
orbital parameters, free of systematic uncertainties.

Concerning the instrumental developments at the DDO: 
Soon, we plan to start using a new CCD system based on a much 
more sensitive detector. While the analysis of the data should
remain the same as described above, we may have to select the
targets more discriminately. In particular, 
it may turn out impractical to observe all 
binaries with periods shorter than one day down to
to the expected limiting magnitude of about 12.5 magnitude. 
Indeed, from the point of astrophysical usefulness, 
it would be advantageous to reduce the deficit of the
intrinsically faint contact systems among spectroscopically
studied binaries of the magnitude-limited sample, 
by attempting to form a volume-limited sample
through giving preference to very short-period systems.

\acknowledgements

While many persons have participated in this program and
have either co-authored the previous papers or their
contributions have been acknowledged there, 
special thanks are due to Dr.\ Hilmar Duerbeck
who contributed to setting the goals of
the program in its early stages when
it was concerned mostly with the center-of-mass velocities for
contact binaries for a planned spatial-velocity investigation.

The author would like to thank Stefan Mochnacki and Mel Blake
for reading and commenting on an early version of the paper. Thanks
are also due to the anonymous referee for two very careful and
constructive reviews.

Support from the Natural Sciences and Engineering Council of Canada
is acknowledged with gratitude. 


\clearpage

\noindent
Captions to figures:

\bigskip

\figcaption[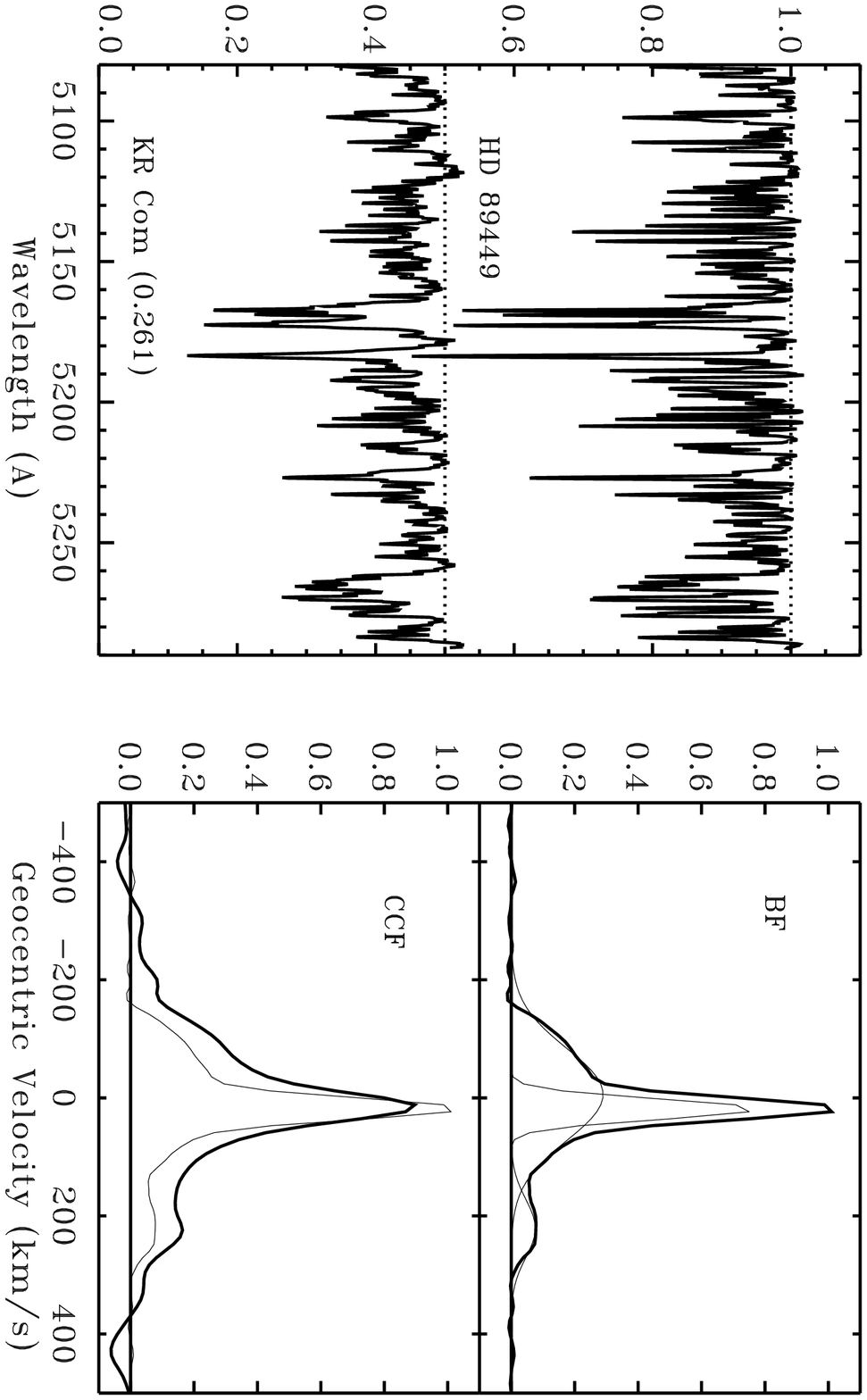] {\label{fig1}
The left panel shows spectra of the sharp-lined template star, 
HD~89449 (upper spectrum), and of the close binary star 
KR~Com (lower spectrum, shifted down by 0.5 
in the observed flux). The spectral types of the stars 
are F6IV and G0IV, respectively. 
The right upper panel shows the broadening function (BF)
obtained by our linear de-convolution using the two spectra in the left 
panel. When measuring radial-velocities of a binary, we initially fit 
the whole triple feature by Gaussians, then subtract the sharp-line 
component and repeat the determination for the close binary.
The three components of the BF are shown by thin lines. 
The lower part of the right panel shows the cross-correlation function
(CCF, thick line), in comparison with the BF (thin line), both obtained
from the same spectra at left. The CCF has much lower resolution than the
BF, but also shows negative excursions in the zero (baseline) level. 
While Gaussians may be a reasonable tool for measurement of
radial velocities from the CCF's, the BF's are much better defined; 
note the much steeper outer ends of the BF relative to the Gaussians.
We discuss extensively the systematic uncertainties of this
type in Section~\ref{syst}.
}

\figcaption[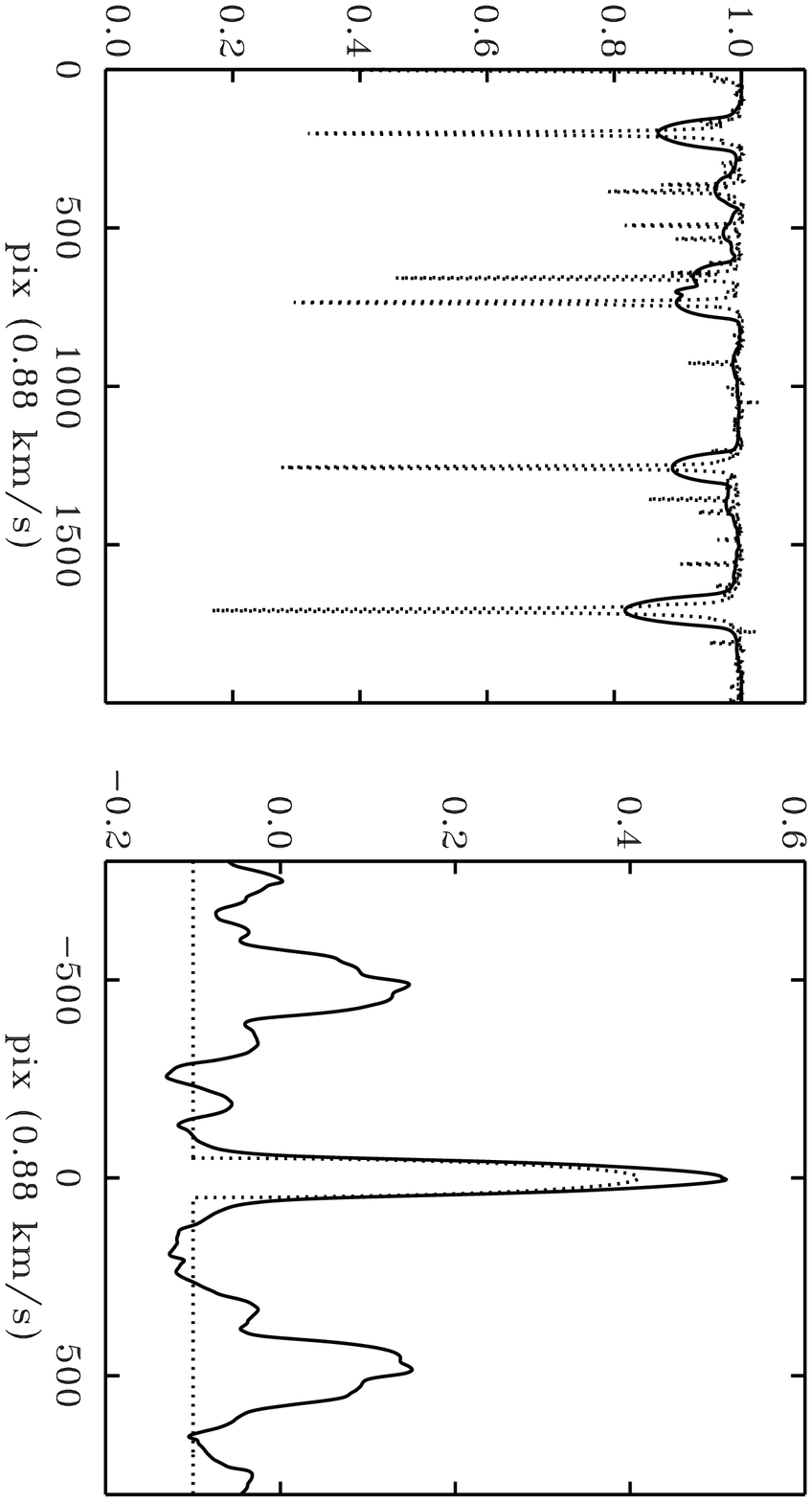] {\label{fig2}
This figure shows an experiment in data processing. 
In the left panel, a high-resolution spectrum, rebinned
to equal velocity steps of 0.88 km~s$^{-1}$, is shown without
any additional broadening (dotted line) and with 
rotational broadening of $V \sin i = 88$ km~s$^{-1}$.
The CCF for the two spectra is shown in the
right panel. Notice the strong positive fringes outside the main
correlation peak as well as the shift of the quasi-baseline
well below the expected zero level; the actual broadening function 
has been shifted down by $-0.1$ units to visualize the most
likely placement of the local baseline in the vicinity of
the correlation peak. 
}

\figcaption[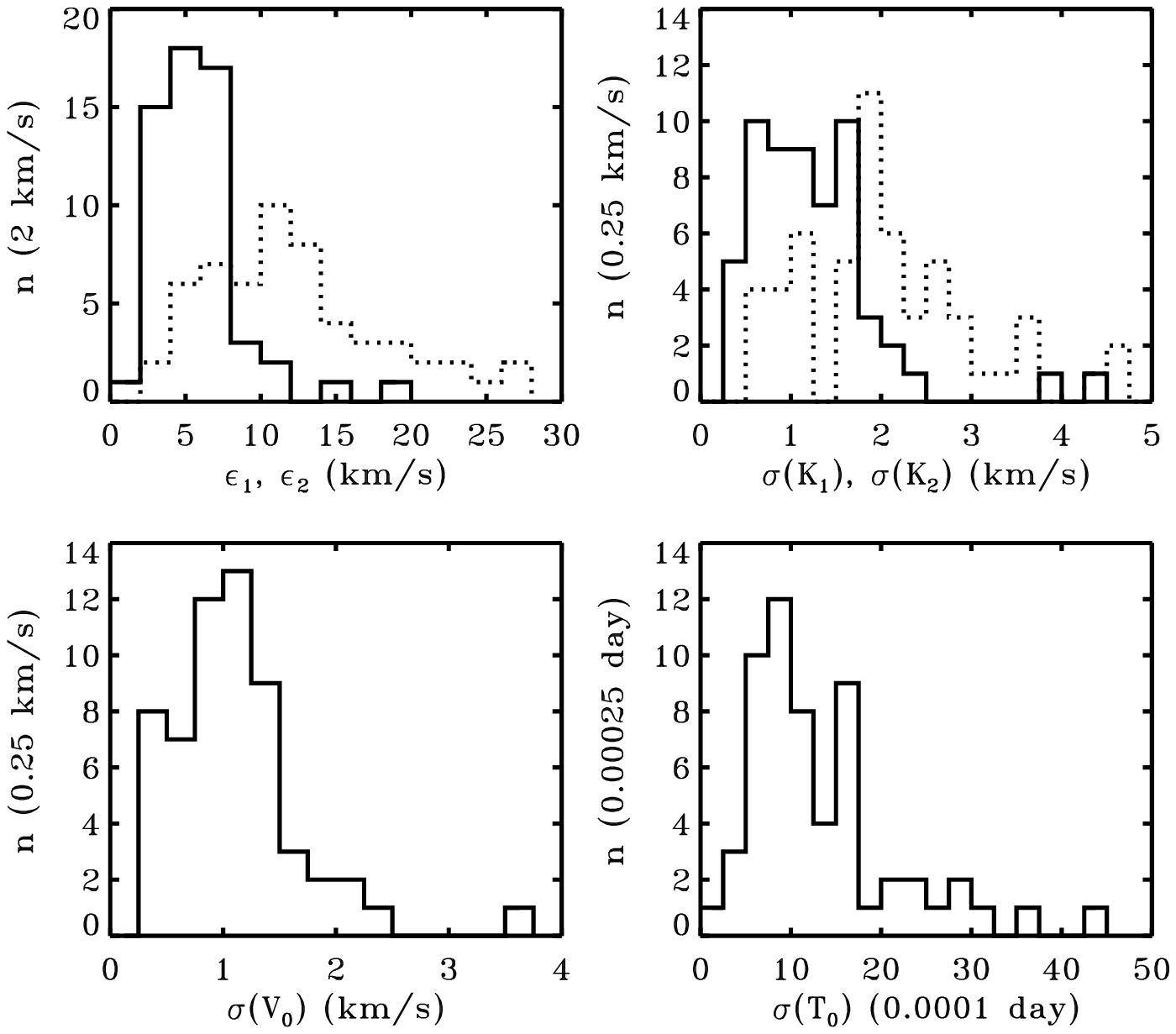] {\label{fig3}
Distributions of mean standard errors for program binaries. The
histograms give the distributions of the error per observation
(for each component), $\epsilon_i$, and of the errors of orbital
parameters $\sigma (V_0)$, $\sigma (K_1)$ and $\sigma (K_2)$, 
all expressed in km~s$^{-1}$. The bin sizes are given in the y-axis
labels. In the two upper panels, full-line histograms are for
the primary components ($\epsilon_1$ and $\sigma (K_1)$)
while the dotted histograms are for the secondary components
($\epsilon_2$ and $\sigma (K_2)$). The last panel gives the
distributions of mean standard errors for the initial epoch $T_0$
in units of 0.0001 days. 
}

\figcaption[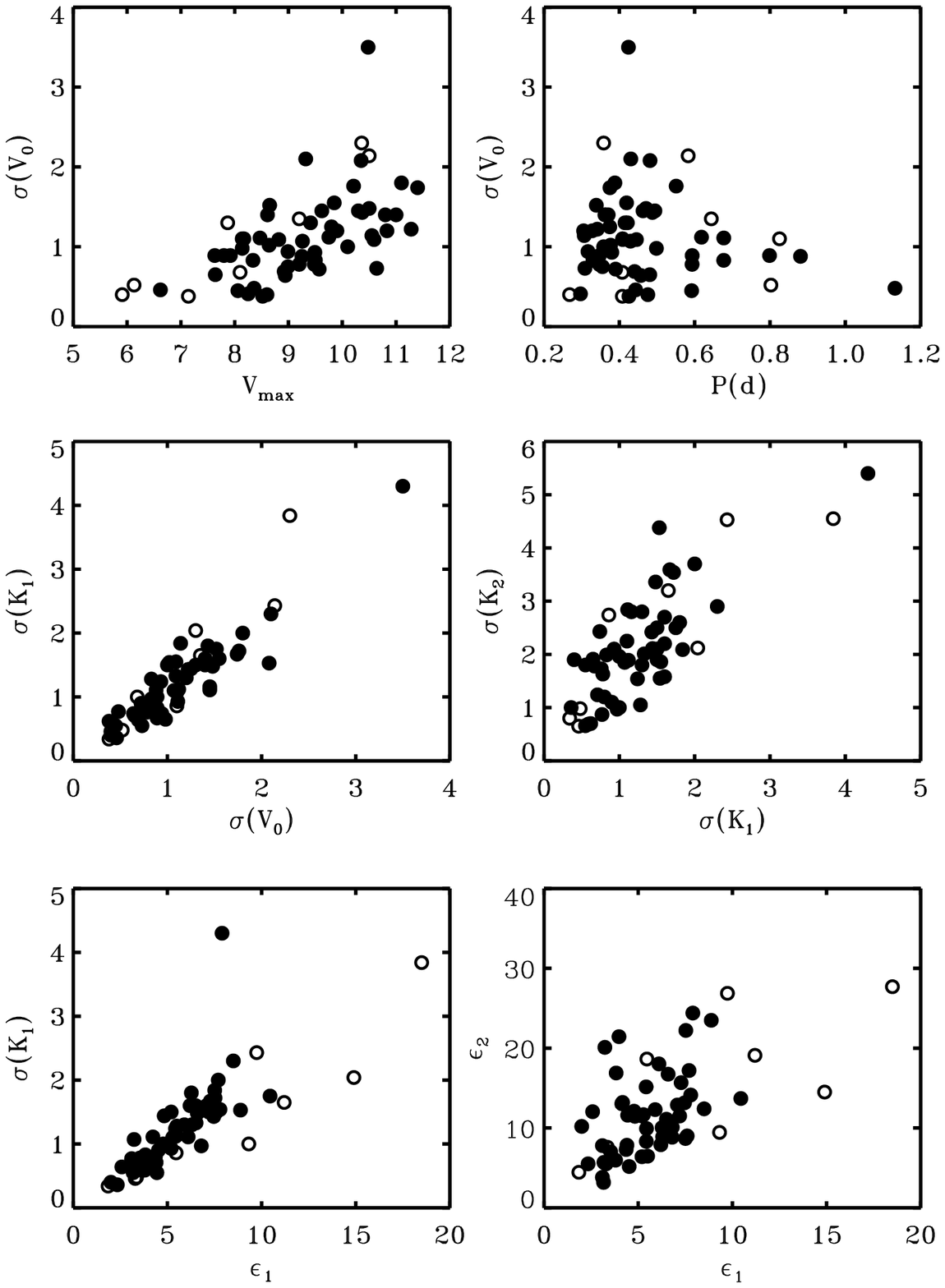] {\label{fig4}
Correlations between various mean standard errors, 
as given in the axis labels.
The two upper panels show $\sigma(V_0)$ as a function of $V_{max}$
and the orbital period, $P$.  $\sigma(V_0)$ is a convenient measure
of the solution quality and correlates tightly with $\sigma(K_1)$,
as shown in the middle left panel. The middle right panel shows
the correlation between  $\sigma(K_1)$ and $\sigma(K_2)$. 
This correlation is
not perfect because of the very large range of mass ratios 
observed among binaries of this program. 
The two lowest panels show the errors
per observation, $\epsilon$; the left panel shows the correlation
between $\epsilon_1$ and $\sigma(K_1)$ while the right panel shows
the correlation between $\epsilon_1$ and $\epsilon_2$. 
In all panels binaries analyzed through subtraction
of the third-component signatures from the broadening functions
are marked by open circles. All quantities
are expressed in km~s$^{-1}$.
}

\figcaption[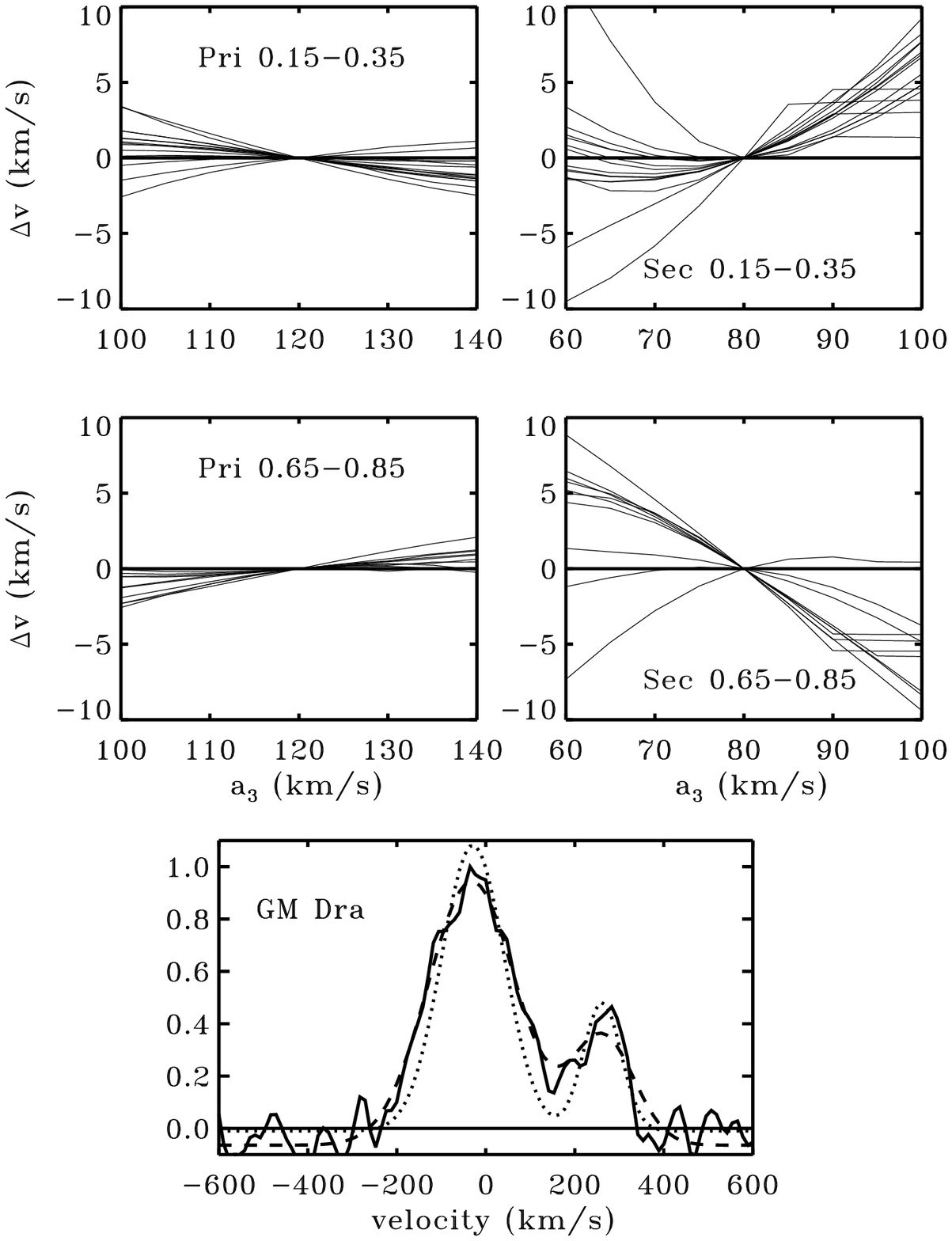] {\label{fig5}
The four upper panels show shifts in the measured velocities for
the primary (left panels) and secondary (right panels)
components of GM~Dra (Paper~VI) versus the Gaussian-width 
parameter $a_{3i}$ (see text). Each line is for one broadening
function at one orbital phase within ranges around
the two orbital quadratures, as marked in the panels.
The lowest single panel shows one broadening function 
of GM~Dra at the orbital phase 0.283, approximated by the
Gaussians with the width parameters [$a_{31}$, $a_{32}$]
considered most extreme for this case:
[100, 60] km~s$^{-1}$ (dotted line) and [140, 100] km~s$^{-1}$
(broken line).
}


\begin{thebibliography}{}

\bibitem[Evans(2000)]{eva00}                                  
    Evans, N. R.
    2000, \aj, 119, 3050
\bibitem[Lu \& Rucinski(1993)]{ruc93a}                       
    Lu, W., \& Rucinski, S. M. 
    1993, \aj, 106, 361
\bibitem[Lu \& Rucinski(1999)]{ddo1}                         
    Lu, W., \& Rucinski, S. M. 
    1999, \aj, 118, 515 (Paper I)
\bibitem[Lu et al.(2001)]{ddo4}                              
    Lu, W., Rucinski, S. M. \& Ogloza, W. 
    2001, \aj, 122, 402  (Paper~IV)
\bibitem[Mochnacki \& Doughty(1972a)]{MD72a}
    Mochnacki, S. W. \& Doughty, N. A.
    1972a, \mnras, 156, 51
\bibitem[Mochnacki \& Doughty(1972b)]{MD72b}
    Mochnacki, S. W. \& Doughty, N. A.
    1972b, \mnras, 156, 243
\bibitem[Morbey(1975)]{mor75}                                
    Morbey, C. 
    1975, \pasp, 87, 689
\bibitem[Press et al.(1992)]{numrec}                   
    Press, W. H., Teukolsky, S. A.
    Vettering, W. T. \& Flannery, B. P.
    1992, Numerical Recipes in Fortran, Second Edition,
    Cambridge Univ.\ Press
\bibitem[Rucinski(1992)]{ruc92}                             
    Rucinski, S. M.
    1992, \aj, 104, 1968
\bibitem[Rucinski(1998)]{ruc98}                             
    Rucinski, S. M.
    1998, \aj, 116, 2998
\bibitem[Rucinski(1999)]{ruc99}                             
    Rucinski, S. M.
    1999, ``Precise Stellar Radial Velocities'', 
    ASP Conf.\ Ser.\ Vol.185, 
    eds. J. B. Hearnshaw \& C. D. Scarfe, p.82
\bibitem[Rucinski \& Lu(1999)]{ddo2}                        
    Rucinski, S. M. \& Lu, W.
    1999, \aj, 118, 2451 (Paper II)
\bibitem[Rucinski \& Lu(2000)]{WCrv}                        
    Rucinski, S. M. \& Lu, W.
    2000, \mnras, 315, 587
\bibitem[Rucinski et al.(1993)]{ruc93b}                     
    Rucinski, S. M., Lu, W. \& Shi, J
    1993, \aj, 106, 1174
\bibitem[Rucinski et al.(2000)]{ddo3}                      
     Rucinski, S. M., Lu, W. \& Mochnacki, S. W.
     2000, \aj, 120, 1133 (Paper III)
\bibitem[Rucinski et al.(2001)]{ddo5}                       
     Rucinski, S. M., Lu, W., Capobianco, C. C. 
     Mochnacki, S. W., Blake, M. 
     2001, \aj, 122, 1974 (Paper~V)
\bibitem[Rucinski et al.(2002)]{ddo6}                       
     Rucinski et al.
     2002, \aj, submitted (Paper~VI)
\bibitem[Stefanik et al.(1999)]{ste99}                      
     Stefanik, R. P., Latham, D. W. \& Torres, G
     1999, ``Precise Stellar Radial Velocities'', 
     ASP Conf.\ Ser.\ Vol.185, 
     eds. J. B. Hearnshaw \& C. D. Scarfe, p.354
\bibitem[Sugars \& Evans(1994)]{SE94}                        
     Sugars, B. J. A \& Evans, N. R.
     1994, \jrasc, 88, 270
\bibitem[Zucker \& Mazeh(1994)]{todcor1}                     
     Zucker, S. \& Mazeh, T.
     1994, \apj, 420, 806
\bibitem[Zucker et al.(1995)]{todcor3}                      
     Zucker, S., Torres, G. \& Mazeh, T.
     1995, \apj, 452, 863

\end{thebibliography}
\end{document}